\newcommand{\CLASSINPUTtoptextmargin}{2.4cm}
\newcommand{\CLASSINPUTbottomtextmargin}{2.56cm}
\documentclass[conference]{IEEEtran}
\IEEEoverridecommandlockouts
\usepackage{cite}
\usepackage{amsmath,amssymb,amsfonts}
\usepackage{algorithmic}
\usepackage{graphicx}
\usepackage{textcomp}
\usepackage{caption}
\usepackage{subcaption}
\usepackage{xcolor}
\def\BibTeX{{\rm B\kern-.05em{\sc i\kern-.025em b}\kern-.08em
    T\kern-.1667em\lower.7ex\hbox{E}\kern-.125emX}}

\usepackage{fancyhdr}

\begin{document}

\title{What-if Analysis Framework for Digital Twins in 6G Wireless Network Management}

\author{
\IEEEauthorblockN{
Elif Ak\IEEEauthorrefmark{1}, 
Berk Canberk\IEEEauthorrefmark{2}\IEEEauthorrefmark{5}, Vishal Sharma\IEEEauthorrefmark{3}, Octavia A. Dobre\IEEEauthorrefmark{4} 
Trung Q. Duong\IEEEauthorrefmark{3}\IEEEauthorrefmark{4}\\
}

\IEEEauthorblockA{
 \IEEEauthorrefmark{1}Department of Computer Engineering, Istanbul Technical University, Istanbul, Turkey}

\IEEEauthorblockA{
 \IEEEauthorrefmark{2}School of Engineering and Built Environment, Edinburgh Napier University, Edinburgh, UK}
 
\IEEEauthorblockA{
 \IEEEauthorrefmark{3}Queen’s University Belfast, BT7 1NN Belfast, UK}
 
\IEEEauthorblockA{
 \IEEEauthorrefmark{4}Memorial University, St. Johns,
NL A1C 5S7, Canada}

\IEEEauthorblockA{
 \IEEEauthorrefmark{5}Department of Artificial Intelligence and Data Engineering, Istanbul Technical University, Istanbul, Turkey}

 	Emails: akeli@itu.edu.tr, b.canberk@napier.ac.uk, v.sharma@qub.ac.uk, odobre@mun.ca, tduong@mun.ca 

}

%\IEEEspecialpapernotice{(Invited Paper)}
\maketitle

\thispagestyle{fancy}   
\fancyhead{}                
\lhead{Accepted by 2024 IEEE The 20th International Wireless Communications \& Mobile Computing Conference (IWCMC), \copyright2024 IEEE}
\cfoot{}

\begingroup\renewcommand\thefootnote{\textsection}
\endgroup
\begin{abstract}
This study explores implementing a digital twin network (DTN) for efficient 6G wireless network management, aligning with the fault, configuration, accounting, performance, and security (FCAPS) model. The DTN architecture comprises the Physical Twin Layer, implemented using NS-3, and the Service Layer, featuring machine learning and reinforcement learning for optimizing carrier sensitivity threshold and transmit power control in wireless networks. We introduce a robust ``What-if Analysis" module, utilizing conditional tabular generative adversarial network for synthetic data generation to mimic various network scenarios. These scenarios assess four network performance metrics: throughput, latency, packet loss, and coverage. Our findings demonstrate the efficiency of the proposed what-if analysis framework in managing complex network conditions, highlighting the importance of the scenario-maker and the impact of twinning intervals on network performance.
\end{abstract}

%\begin{IEEEkeywords}
%Digital Twin, CTGAN, Wireless Networks, What-%if Analysis, Digital Twin Networks 

%\end{IEEEkeywords}

\section{Introduction}
In the ever-evolving domain of network technology and communication infrastructure, network management is a pivotal and continuously advancing area of research. Despite the ongoing advancements, the ISO Telecommunications Management Network's broader, well-known model for network management, namely fault, configuration, accounting, performance, and security (FCAPS), remains a foundational framework in this realm~\cite{isoISOIEC749841989}.

One of the primary challenges in 6G wireless network management is the ability to proactively identify and mitigate issues before they impact the network's performance or user experience. While effective in reactive scenarios, traditional network management approaches often fall short in predictive capabilities and dealing with the complexities of modern wireless, backbone or datacenter networks. This limitation becomes particularly evident in the context of the FCAPS-aided 6G networks \cite{taleb20226g}, where each area demands real-time monitoring and response, foresight, and preemptive planning. 

\begin{figure*}[t!]
    \centering
    \vspace*{0.05in}
    \includegraphics[width=.85\linewidth]{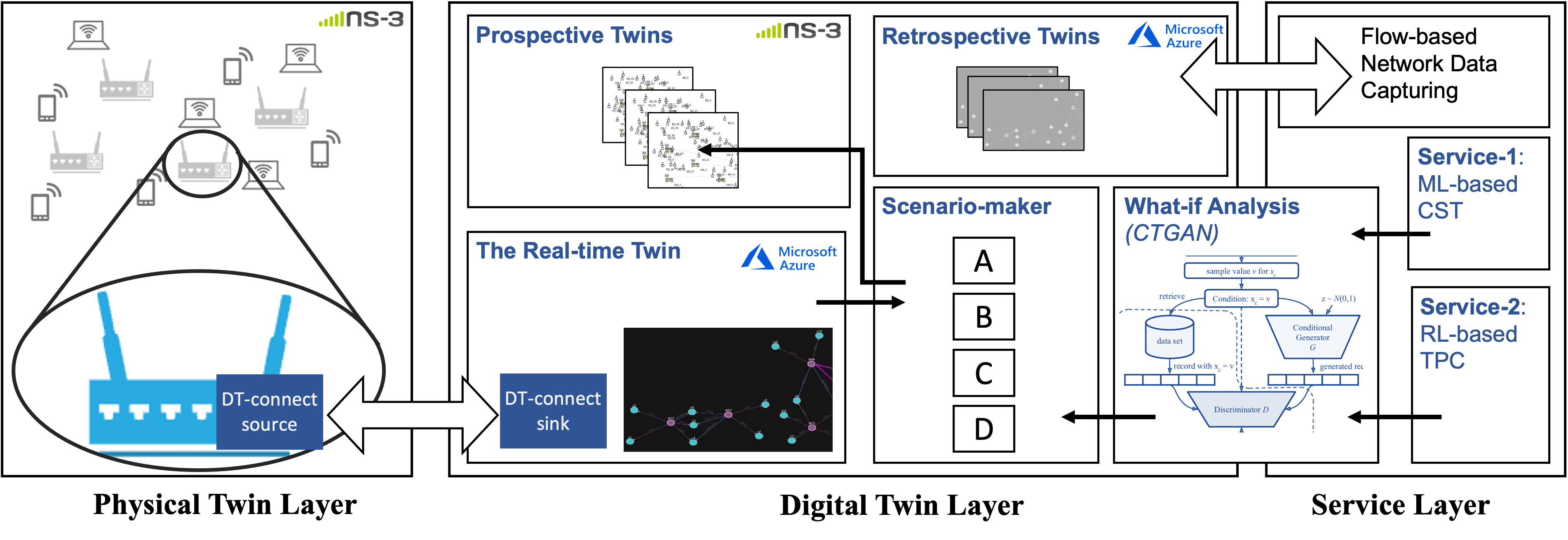}
    \caption{Three-Layered Architecture: What-if Analysis Framework for Digital Twins in 6G Wireless Network Management.}
    \label{fig:dtn}
\end{figure*}

Digital twin networks (DTNs), with their capability to create a virtual mirror of physical network infrastructures, align closely with the principles of the FCAPS 6G network management. They offer an interactive platform that significantly enhances network management across all five facets by allowing real-time monitoring, management, and optimization of network functions. However, the DTN approach does not have the capabilities to meet all five FCAPS criteria as is and must be carefully %designed, developed and 
deployed across the network. According to insights from Autodesk Inc.\footnote{https://www.autodesk.com/ [Last Accessed: November 2023]}, a front-runner in DTN research, digital twins can be classified into five levels based on their autonomy and sophistication. The fourth level, known as the Comprehensive Twin, is notable for its integration of ``what-if" analysis capabilities. This level leverages advanced modelling, simulation of potential future scenarios, and prescriptive analytics, making it a critical point in aligning with the FCAPS model. %It is the first level that sufficiently supports the entirety of the FCAPS criteria, just preceding the fifth and most advanced level. In other words, it is clear to say that having a well-designed ``what-if" approach in DTNs also accelerates the FCAPS principles in network management.

In this context, our study introduces a ``what-if" analysis approach, utilizing a conditional tabular generative adversarial network (CTGAN) to address critical aspects of the FCAPS model — specifically, Fault, Configuration, and Performance. While current DTN research underscores the importance of ``what-if" analysis in network management, comprehensive frameworks supporting the FCAPS model are inadequate \cite{RAY2021102180}. 
Therefore, our framework aims to bridge the existing gaps in DTN applications, tailoring them to meet the requirements of contemporary network management. \color{black} This approach, initially developed and tested in a Wi-Fi 6 environment, holds significant potential for the emerging 6G ecosystem. The Wireless Local Area Networks (WLANs) use case, though specific to Wi-Fi 6, provides a solid foundation for 6G network management strategies. 
The main contributions of the proposed approach are as follows:
\begin{itemize}
    \item  Implementing a DTN architecture for wireless network management that aligns with the FCAPS 6G networks.
    \item Introducing a what-if analysis framework using CTGAN for generating synthetic data across various scenarios.
    \item Employing scenario-maker module in digital twin layer to assess key network performance metrics and demonstrate the impact of intelligent configuration decisions.
    \item Exploring the effect of twinning intervals on overall network performance by emphasizing the significance of DTN on network management.
\end{itemize}

\section{Related Works}
Recent advancements in network management have seen a growing focus on DTNs \cite{10148936}, recognized for their potential to enhance network performance and align with the FCAPS model \cite{9795043}. Prior studies have explored the integration of machine learning and reinforcement learning techniques for network optimization, particularly in the context of adjusting network parameters \cite{9839640}, like transmit power and sensitivity thresholds in WLANs. 

In the realm of 6G network management, digital twins have been explored across various facets, such as network slicing \cite{9310275}, improving real-time data processing \cite{9170905}, reducing offloading latency \cite{9174795}, and open 6G radio access networks \cite{10179151}. These studies collectively demonstrate that the DTN approach significantly enhances performance across diverse 6G network metrics. However, as highlighted in a recent comprehensive survey \cite{9923927}, all 6G DTN proposals, including resource allocation and network slicing methods, need to undergo thorough testing via what-if scenarios. This step is crucial to prevent any adverse impacts on the physical network. In other words, the role of what-if analysis in DTNs is crucial for envisioning future scenarios based on predicted configurations \cite{9839640}. While some research has explored the behaviour of networks using different degrees of freedom to define KPIs and configurations, these often lead to increased computational time due to sequential testing of all possible configurations \cite{9468224}. Other works have proposed case studies in backbone networks, assessing load balancing in protocols like BGP, but have not extensively covered the long-term impact of new configurations \cite{10154393}.

Several studies have focused on using synthetic data for digital twin networks in 6G, emphasizing the transfer learning capabilities and addressing the challenges in various networking scenarios without synthetic data \cite{10198573}. The current literature highlights the importance of data-intensive simulations in what-if analysis \cite{9468224, 9795043}. Models such as the Conditional Tabular Generative Adversarial Network (CTGAN) \cite{xu2019modeling} are particularly promising for generating synthetic data to predict future network behaviours, including mobility patterns and data traffic dynamics.

Although existing literature extensively explores the potential of DTNs in network management, there is a noticeable gap in effectively combining What-if analysis with the Service Layer of DTNs for more dynamic and efficient optimization of network parameters. Our contributions specifically address this gap by integrating CTGANs for predictive scenario generation and applying optimization strategies. %for wireless configurations. %This approach enhances computational efficiency and provides more comprehensive impacts under varied conditions.

\section{What-if Analysis Framework for Digital Twins}
Our proposed What-if Analysis Framework for Digital Twins in 6G wireless network management presents a sophisticated, three-layered architecture designed to optimize network management in dense wireless environments. This framework comprises the Physical Twin Layer, the Digital Twin Layer, and the Service Layer, each playing a distinct yet interconnected role as shown in Fig. \ref{fig:dtn}. 
\subsection{Physical Twin Layer}
Our focus is on densely deployed wireless scenarios characterized by overlapping coverage areas of multiple Base Stations (BSs), forming what is known as Overlapping Basic Service Sets (OBSS). In our model, all BSs are configured to operate on the same channel within the 2.4 GHz band. This choice intentionally models overlapped scenarios and emphasizes the study of Carrier Sensitivity Threshold (CST) and Transmit Power Control (TPC) configurations on the same channel rather than balancing across different channels. The DT-Connect \textit{Source} module in each BS collects load metrics through the agent program. These agent programs are crucial for gathering comprehensive data, including BS configurations, client details, traffic information, and packet logs. The collected data is then sent to the DT-Connect \textit{Sink} module to be processed in the Digital Twin Layer.

The Physical Twin Layer of our DTN is conceptualized using a network simulation environment, specifically through the NS-3 simulation. This simulation is a practical replica of a physical network, enabling us to analyze and manage network configurations in a controlled yet realistic setting. 
It mimics the complexities of a dense 6G wireless network and provides the necessary data infrastructure to support advanced ``what-if" analyses in the subsequent Service Layer of the DTN.

\subsection{Digital Twin Layer}
In developing the Digital Twin Layer for the network management framework, we have integrated Microsoft Azure's Digital Twin cloud services. This integration effectively simulates and analyzes the network scenarios, particularly focusing on the CST and TPC in our wireless model. There are four parts in this layer, as explained in detail below.

\subsubsection{\textbf{DT-Connect Source and Sink Flow}}
Microsoft Azure IoT Hub serves as the gateway connecting our Physical Twin Layer, simulated through NS-3, to the Digital Twin Layer. Agent programs installed on the simulated BSs relay information to their corresponding IoT Hub instances. This setup ensures a seamless flow of data from the physical network layer to the digital twin.

\subsubsection{\textbf{Digital Twins}} The Digital Twin module, within the Digital Twin Layer, consists of three distinct and interconnected sub-modules: The Real-time Twin, Retrospective Twins, and Prospective Twins as follows:

\begin{itemize}
    \item \textit{The Real-time Twin} is developed using Microsoft Azure Digital Twins (ADT), providing a live, real-time digital representation of the physical network. It mirrors the current state of the network, including the configurations, performance, and interactions within the network, using models coded in Digital Twins Definition Language (DTDL). It includes key parameters of CST and TPC, such as bandwidth usage, signal strength, and node connectivity. In our ADT implementation, we have defined two primary interfaces for BSs and User Equipments (UEs), separately. These interfaces encapsulate essential aspects of the wireless network. \textit{Property Fields} of DTDL represents the status of physical objects. For BS interfaces, we store information such as SSID and Channel. In the case of UEs, received and transmitted packet counts are recorded through the connected BS of the UEs. \textit{Telemetry Data} of DTDL includes real-time measurements that are crucial for dynamic network analysis but are not stored as a permanent part of the digital twin. For instance, BS telemetry includes CPU utilization, while UE telemetry encompasses received signal strength and associated BSs' MAC addresses. We have established relationships between BS and client models, directly correlating to the packet logs sensed by BSs. These relationships include data like the last timestamp of signals detected by the BSs. 
    \item \textit{Retrospective Twins} are also modeled using Microsoft ADT. They represent the historical states of the network, providing insights into past performance, configurations and network status. The graph-based historical representation also allows relations between network nodes to be maintained naturally. 
    \item  \textit{Prospective Twins} are generated from the scenario-maker module and represent potential future states of the network. They are modelled using NS-3 logs, offering a forward-looking perspective by simulating various ``what-if" scenarios. These twins allow us to experiment with different network configurations and settings, i.e. varying CST and TPC, to predict their potential impact.
\end{itemize}

\subsubsection{\textbf{Scenario-maker}} This module, an integral component of our Digital Twin Layer design, plays a pivotal role in testing various ``what-if" network scenarios, each designed to test different aspects of network behaviour and resilience. This module is adeptly configured to generate and assess \textit{four distinct scenarios}, leveraging both real-time data and synthetic data generation techniques as follows.
\begin{itemize}
    \item \textit{Scenario-A: Existing Behavior} continues to replicate the current behaviour of the network. It utilizes live-stream data sourced directly from \textit{The Real-time Twin}, providing an up-to-date and accurate representation of the existing network status. This scenario serves as a baseline, assessing the near-future or short-term impacts of current network configurations and user interactions within the DTN framework. In other words, Scenario-A carried out in this paper is the scenario presented in the Experimental Setup (Section IV-A) section, which follows the initial setup during the experiment, that is, maintaining the same topology array with exactly the same number of BSs and the same number of UEs.
    \item \textit{Scenario-B: High User Density} modelling addresses the challenges of simulating the network's response to a continuous increase in user numbers, progressively rising to 50\% \textit{(number of users grows steadily one by one from the beginning of the scenario until it reaches a 50\% increase)}. This scenario mimics the future case to understand the network's capability to handle peak usage times and increased user demand, testing the scalability and adaptability of the network infrastructure.

    \item \textit{Scenario-C: Traffic Rate Variations} contains three sub-scenarios (C-1, C-2, and C-3) to simulate varying levels of network traffic increases — 20\%, 40\%, and variable loads according to time-of-day patterns, respectively. These sub-scenarios assess how the network copes with different degrees of traffic load, especially during varying time periods, such as low usage during late-night hours versus high usage during the day.

    \item \textit{Scenario-D: Unforeseen Circumstances} tests the network's robustness under unforeseen or extreme conditions, including sudden traffic spikes and a massive influx of new connections. The difference between Scenario-D from Scenario-B and Scenario-C is that it focuses on the network's preparedness for unexpected events with sudden increases (instead of gradually), ensuring that it remains resilient and functional under stress.
\end{itemize}
%EA: add CTGAN arch figure and explanation here
Scenarios B, C, and D employ a CTGAN model to create realistic yet hypothetical data sets that mimic new mobile users, connections, and varied traffic patterns. CTGAN is adept at handling heterogeneous tabular data comprising both numeric and categorical features using a combination of mode-specific normalization for numeric features and a conditional generator for categorical features. This methodology allows CTGAN to effectively mimic real-world data distributions, making it possible to generate realistic synthetic data that reflects the future behaviour of new mobile users, connections, and varied traffic patterns.

This includes a generator network $G$ that learns to generate synthetic data samples from a noise distribution, while the discriminator $D$ learns to differentiate between real and generated samples. The objective function, Eq. \ref{eq:ctgan} \cite{xu2019modeling}, is as follows: 

\begin{equation}
  \label{eq:ctgan}
  \begin{gathered}[b]
\min_{G} \max_{D} V(D, G) = \mathbb{E}_{x \sim p_{data}(x)}[\log D(x)]  \\
+ \mathbb{E}_{z \sim p_{z}(z)}[\log(1 - D(G(z)))]
  \end{gathered}
\end{equation}

In this formula, $x$ represents real data samples, and $z$ represents points in the generator's input noise space. The expectation $E$ is taken over the real data distribution $p_{data}(x)$ and the generator's noise distribution $p_z(z)$. And, $\mathbb{E}_{x \sim p_{data}(x)}[\log D(x)]$ calculates the expectation of the discriminator's output (probability of real data being classified as real) over the real data distribution. $\mathbb{E}_{z \sim p_{z}(z)}[\log(1 - D(G(z)))]$ calculates the expectation of the discriminator's output (probability of generated data being classified as fake) over the generated data.

\subsubsection{\textbf{What-if Analysis}}
This module utilizes four key performance indicators (KPIs): throughput, $t$, latency, $l$, packet loss, $pl$, and coverage, $c$. For each simulated scenario, these KPIs are calculated and normalized ($\tilde{\mathcal{M}_t}$, $\tilde{\mathcal{M}_l}$, $\tilde{\mathcal{M}_pl}$ and $\tilde{\mathcal{M}_c}$) to ensure consistency in measurement scales. We assign specific weights to each KPI based on their relative importance ($w_t$, $w_l$, $w_{pl}$, $w_c$), which allows us to derive a composite score for each scenario ($CS_A, CS_B, CS_C, CS_D$) as seen in the Eq. \ref{eq:cs}. 
\begin{equation}
  \label{eq:cs}
  \begin{gathered}[b]
    CS_{\text{\{A,B,C,D\}}} = w_{\text{t}}\text{$\tilde{\mathcal{M}_t}$ } + w_{\text{l}} \text{$\tilde{\mathcal{M}_l}$ } + w_{\text{pl}} \text{$\tilde{\mathcal{M}_{pl}}$ } + w_{\text{c}} \text{$\tilde{\mathcal{M}_c}$ }
  \end{gathered}
\end{equation}

Furthermore, each scenario is also weighted ($w_{A}, w_{B}, w_{C}$ and $w_{D}$) to reflect its significance in the overall network assessment, called \textit{Effectiveness} score, $\xi$, as follows: 

\begin{equation}
    \xi = w_{A}CS_A + w_{B}CS_B + w_{C}CS_C + w_{D}CS_D
\end{equation}
In this way, it enables a comprehensive evaluation of the selected network configurations of TPC and CST proposed by the Service Layer, providing insights into their effectiveness under different conditions and scenarios.

\subsection{Service Layer}
The Service Layer in the DTN framework begins with the \textit{Network Flow Preprocessing} module, which prepares and cleans data for analysis. We then selected two specific services for our wireless case study based on our previous two studies \cite{ccakir2022dtwn, 9322153}, focusing on configurations that could potentially conflict. These configurations are chosen with care to avoid packet loss while increasing the coverage. Due to the wide range of possible configurations, intelligent decision-making is crucial. ``Service-1" utilizes Neural Networks (NN) to optimize the CST \cite{9322153}, improving network sensitivity and efficiency. ``Service-2" employs reinforcement learning for adaptive control of TPC \cite{ccakir2022dtwn}, balancing optimal signal strength and coverage against interference and power consumption. After determining new configurations using current and historical data, these are simulated in the DTN using the ``What-if Analysis" module, as previously explained.

\section{Performance Evaluation}
%\subsection{Experiment Setup}
The simulation parameters used in that study can be found in Table \ref{table:sim}. 
As explained in the experiments and results, the number of BSs is changed to 3,9,27,81, and 243 to change the size of the wireless topology. Also, each wireless topology is tested 100 times with a random number of UEs in each BS, changing from 10 to 15. Other details used in the experiments are explained below.

\begin{table}[h]
\centering
\caption{Simulation parameters}
\label{table:sim}
\begin{tabular}{|c|c|}
\hline
Parameter & Value\\
\hline \hline
Scenario & Outdoor \cite{merlin2015tgax}\\ \hline
Channel Band / Bandwidth & 2.4Ghz / 20Mhz\\ \hline
BS/UE Tx Power & 20/15 dBm \\ \hline
Traffic & CBR, HTTP, Video \\ \hline
Traffic per BSS (Mbps) & 100 \\ \hline
%Number of antennas & SISO \\ \hline
RTS/CTS & Disabled \\ \hline
Packet & 1464 bytes \\ \hline
Beacon Interval & 102.4 ms \\ \hline
Guard Interval Duration & 1.6 us\\ \hline
Modulation & 256-QAM \\
\hline
\end{tabular}
\end{table}

\textbf{Topology size effect on running time of what-if analysis}: 
In the initial stage of our experiments, we expanded the size of the wireless topology within the Physical Twin Layer exponentially. This was done to evaluate the runtime of the ``What-if Analysis" and ``Scenario-Maker" modules. In the Service Layer, we examined six different approaches to understand how decision-making impacts the duration of what-if analysis as the topology size increases.

\begin{figure}
     \centering
     \begin{subfigure}[b]{0.49\textwidth}
         \centering
         \includegraphics[width=.85\textwidth]{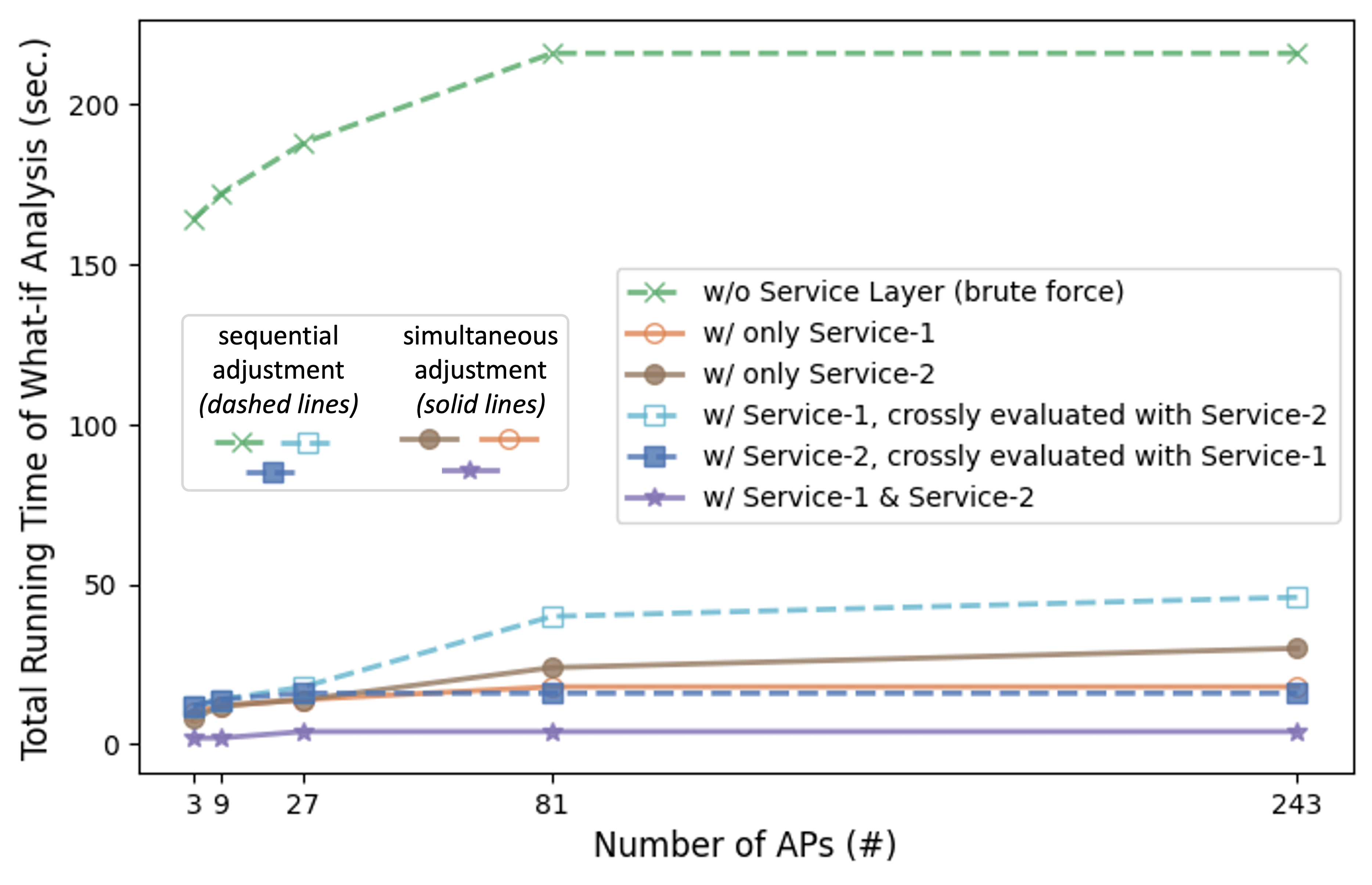}
         \caption{Total running time with changing topology sizes.}
         \label{fig:running_time}
     \end{subfigure}
     \hfill
     \begin{subfigure}[b]{0.49\textwidth}
         \centering
         \includegraphics[width=.9\textwidth]{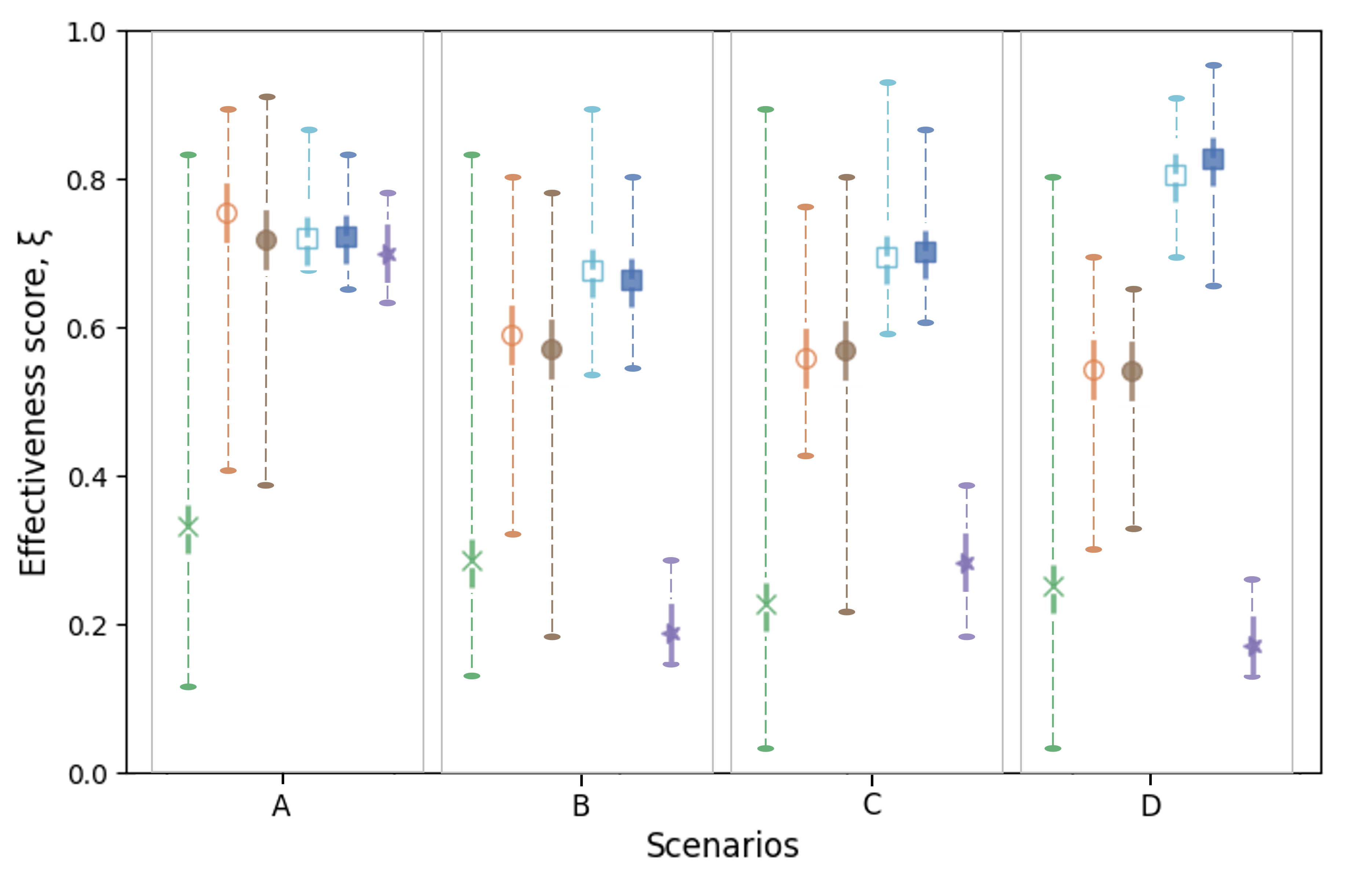}
         \caption{Performance of service layers through scenarios.}
         \label{fig:scenarious}
     \end{subfigure}
        \caption{Comparison of different Service Layer decisions.}
        \label{fig:service_layer}
\end{figure}

The first approach, termed \textit{``without (w/o) Service Layer (brute force)"}, involved testing every possible configuration for TPC and CST. The next two approaches focused separately on Service-1 (CST) and Service-2 (TPC), referred to as \textit{``with (w/) only Service-1"} and \textit{``with (w/) only Service-2"} in Fig. \ref{fig:running_time}. Additionally, we conducted cross-evaluations: using Service-1 for intelligent CST configuration while sequentially testing each TPC choice, labelled as \textit{``with Service-1, crossly evaluated with Service-2"}, and the reverse for \textit{``Service-2, crossly evaluated with Service-1"}. Lastly, we included an approach where Service-1 and Service-2 run their trained models independently to determine specific CST and TPC values, shown as \textit{``w/Service-1 \& Service-2"}.

As expected, sequentially evaluating all potential configurations for a given wireless status becomes increasingly time-consuming, especially as the topology size and the range of CST and TPC possibilities grow. The approach with the shortest runtime was where Service-1 and Service-2 independently ran machine learning and reinforcement learning models to determine TPC and CST values. This result, along with others, is depicted in Fig. \ref{fig:running_time}. These findings also underscore that, without a DT framework, conducting what-if analysis on all possible wireless configuration parameters is impractical in network management.

\textbf{Performance of service layer through what-if scenarios}: We then evaluate the output of the \textit{Scenario-maker} module to understand the performance of the above six Service Layer approaches. These assessments were performed across various wireless topology sizes, each executed 100 times and displayed with minimum, maximum and mean values as a box plot chart in Fig. \ref{fig:scenarious}. We evaluated the performance through \textit{Effectiveness score}, $\xi$, on each of the four scenarios. This measurement, $\xi$, is derived from the ``What-if Analysis" module in the Digital Twin Layer, rather than being a direct output from the actual wireless topology in the Physical Twin Layer. The goal here was to evaluate the performance of the Service Layer approaches prior to implementing the selected configurations in the Physical Twin Layer. Here, the findings reveal that although running Service-1 and Service-2 independently in Digital Twin might minimize runtime, it leads to lower performance across all scenarios (causing 63\% less $\xi$ score), shown as a purple line with a star ($\star$) symbol in Fig. \ref{fig:scenarious}. This outcome is primarily attributed to potential conflicts when these services (Service-1 and Service-2 in our case) are run independently and simultaneously. The most consistent performance was achieved by intelligently calculating one service's output (either Service-1 or Service-2) and then sequentially cross-evaluating all possible configurations for the other service (lines in which light blue with an empty box symbol and dark blue with a filled box symbol, respectively). This approach aligns with the inherent challenges of using two potentially conflicting services in the Service Layer of DTNs for wireless network management.

\begin{figure}
     \centering
     \begin{subfigure}[b]{0.45\textwidth}
         \centering
         \includegraphics[width=.9\textwidth]{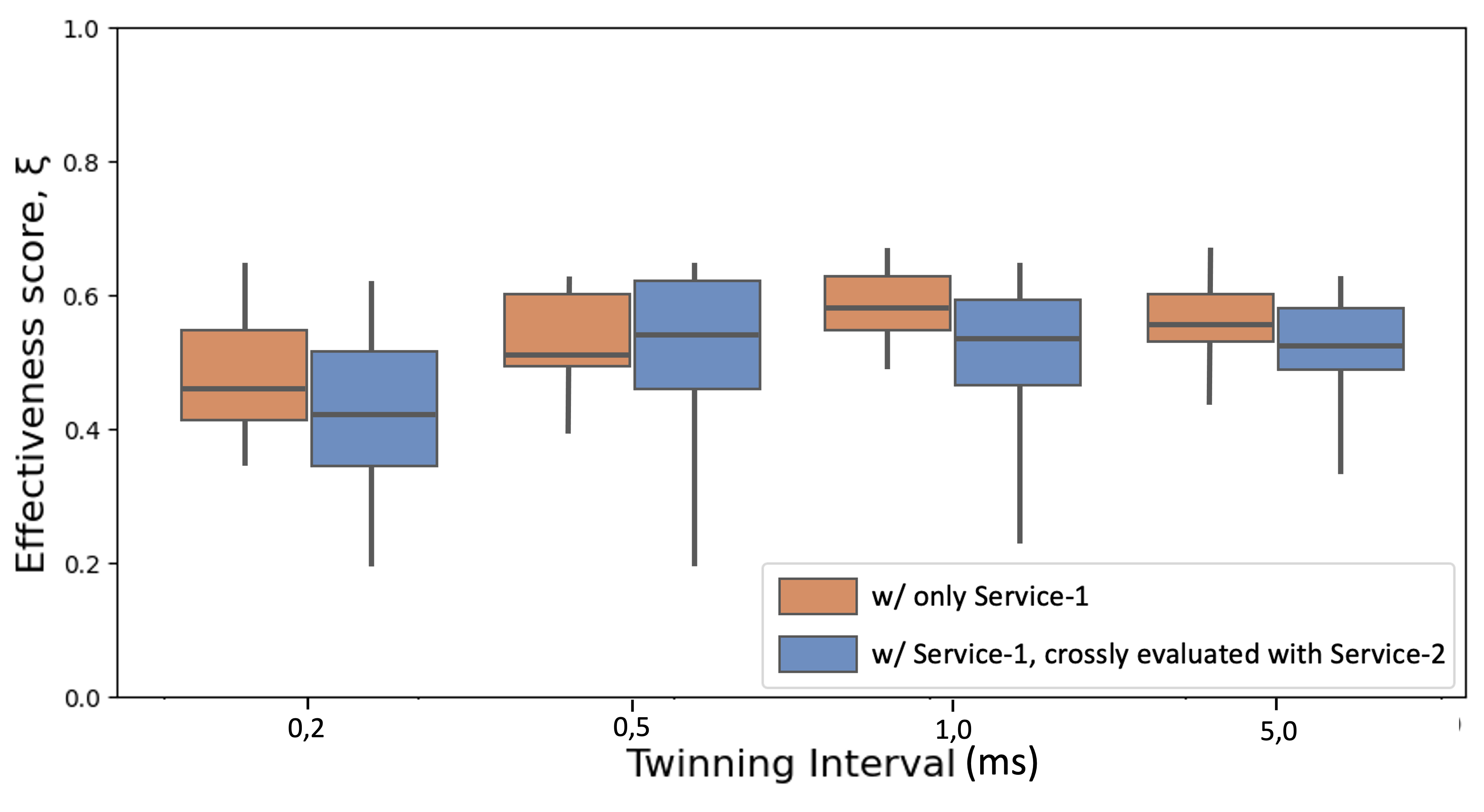}
         \caption{Scenario-A}
         \label{fig:twinning_sec_a}
     \end{subfigure}
     \hfill
     \begin{subfigure}[b]{0.45\textwidth}
         \centering
         \includegraphics[width=.9\textwidth]{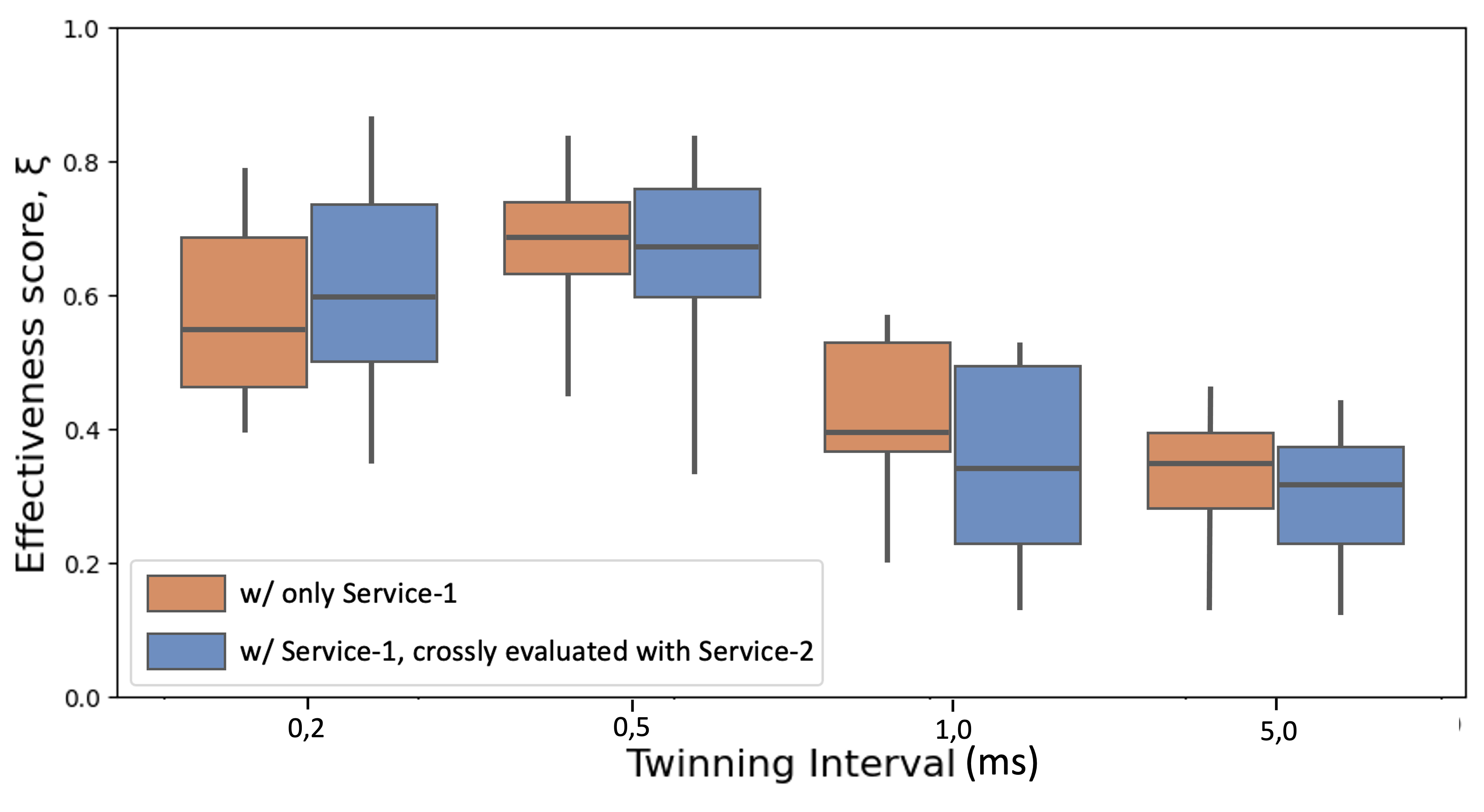}
         \caption{Scenario-All}
         \label{fig:twinning_sec_all}
     \end{subfigure}
        \caption{Twinning interval and performance implications.}
        \label{fig:twinning_rate}
\end{figure}

\textbf{Twinning interval and performance implications}:
In our final experiment, we assessed the performance of the DTN by varying the twinning intervals via Fig. \ref{fig:twinning_rate}. This interval determines the frequency at which data is transferred from the physical network to the digital twin. For this part of the experiment, we applied CST and TPC configurations that achieved an $\xi$ score greater than $0.8$, as determined in the What-if Analysis and Scenario-Maker modules within our proposed three-layered architecture. In the Service Layer, we focused on two approaches: ``with only Service-1" and ``with Service-1, crossly evaluated with Service-2". This choice was based on previous findings, as discussed in Fig. \ref{fig:scenarious}, which demonstrated that using one service or combining two services (one calculated with ML and the other evaluated sequentially) resulted in similar average $\xi$ scores, yet with different confidence intervals across 100 experiments.

Initially, we observed the $\xi$ score in the Physical Twin Layer using only Scenario-A in the Scenario-Maker module, as seen in Fig. \ref{fig:twinning_sec_a}. Subsequently, we also incorporated all scenarios to measure the $\xi$ score from the real wireless topology, selecting configurations that achieved an $\xi$ score greater than 0.8 across all four scenarios.  The outcomes, depicted in Fig. \ref{fig:twinning_sec_a}, compare the performance when only Scenario-A (Existing Behavior) is used against including all scenarios in the what-if analysis. The results, as shown in Fig. (\ref{fig:twinning_sec_all}), indicate that a diverse range of scenarios in the analysis improves the chances of selecting optimal network configurations, aligning with FCAPS objectives in 5G/6G networks.

Furthermore, using just one service in the Service Layer for new configuration settings resulted in a narrower confidence interval but a lower mean $\xi$  score in the real wireless topology. Conversely, considering two different services led to higher mean $\xi$  scores but a larger confidence interval. Quantitatively, the comprehensive integration of all scenarios through our CTGAN-based Digital Twin methodology facilitates a 43\% enhancement in $\xi$ score, but at the expense of a 12\% expansion in the confidence interval.  Additionally, a longer twinning interval was found to reduce performance, as it can lead to missed transient peaks and loss of crucial network traffic details by averaging data over extended periods. Thus, the twinning interval is a critical parameter that must be tuned according to the specific network topology. 

\section{Conclusion}
This work highlights the potential of digital twin networks in enhancing 6G wireless management. By integrating advanced simulation techniques and analytical models, digital twin networks effectively address key FCAPS areas. The utilization of CTGAN for synthetic data generation in ``What-if Analysis" significantly contributes to understanding network behaviour under various hypothetical scenarios. The study underscores the critical role of twinning intervals and scenario diversity in achieving optimal network configurations. Overall, this approach offers a promising direction for future developments in 5G/6G management, leveraging digital twins for more resilient, efficient, and adaptable network infrastructures.

\section*{Acknowledgements}
The work of B. Canberk is supported by The Scientific and Technological Research Council of Turkey (TUBITAK) 1515 Frontier R\&D Laboratories Support Program for BTS Advanced AI Hub: BTS Autonomous Networks and Data Innovation Lab, Project 5239903. The work of V. Sharma was supported by the UK Department for Science, Innovation and Technology under the Future Open Networks Research Challenge project TUDOR (Towards Ubiquitous 3D Open Resilient Network). The work of O. A. Dobre was supported in part by the Canada Research Chairs Program CRC-2022-00187. The work of T. Q. Duong was supported in part by the Canada Excellence Research Chair (CERC) Program CERC-2022-00109.

\bibliographystyle{IEEEtran}
\bibliography{main.bib}

\end{document}